\documentclass{WileyChemistry-template}

\title{\raggedright Structures and stabilities of mixed clusters of fullerene and coronene molecules}

\author{
\begin{minipage}{\textwidth}
	Naemi Florin,*\textsuperscript{[a]} Henning Zettergren,\textsuperscript{[a]} and Michael Gatchell\textsuperscript{[a]}
\end{minipage}
}

\newcommand{\affiliation}{
\begin{itemize}

\item[{[a]}] Naemi Florin*, Prof.\ Henning Zettergren, Dr.\ Michael Gatchell\\
Department of Physics, Stockholm University, 106 91 Stockholm, Sweden\\
E-mail: naemi.florin@fysik.su.se
\end{itemize}
}

\renewcommand{\dedication}{
	\begin{minipage}{\textwidth}
	\end{minipage}
}

\renewcommand{\abstract}{We have performed molecular dynamics simulations on the formation of mixed molecular clusters of buckminsterfullerene and coronene, $(\mathrm{C}_{24}\mathrm{H}_{12})_n(\mathrm{C}_{60})_{N-n}$. We report on our findings on the structures and their relative stabilities for cluster sizes $N=5$ and 13 and for all possible combinations of the two species within these sizes, including the pure clusters of each type. Generally, we see that the two species mix rather poorly and that compactly bound clusters are favoured over spatially extended ones. For a given ratio of coronene and fullerene, clusters with one or two coronene stacks tend to be more stable than those with a larger number of stacks. In the case of small clusters, the coronene and fullerene molecules tend to separate into two different cluster parts. For larger clusters, this is often but not always the case.

}

\newcommand{\keywords}{
	Fullerenes \textbullet\ 
	PAHs \textbullet\ 
	Structures \textbullet\ 
	Clusters \textbullet\ 
	Molecular Dynamics
}

\begin{document}
%%%%%%%%%%%%%%%%%%%%%%%%%%%%%%%%%%%%%%%%%%%%%%%%%%%%%%%%%%
%%%%%%%%%%%%%%%%%%%%%%%%%%%%%%%%%%%%%%%%%%%%%%%%%%%%%%%%%%
%%%%%%%%%%%%%%%%%%%%%%%%%%%%%%%%%%%%%%%%%%%%%%%%%%%%%%%%%%

\twocolumn[\vspace{-1.5cm}\maketitle\vspace{-1cm}
	\textit{\dedication}\vspace{0.4cm}]
\small{\begin{shaded}
		\noindent\abstract
	\end{shaded}
}

\begin{figure} [!b]
\begin{minipage}[t]{\columnwidth}{\rule{\columnwidth}{1pt}\footnotesize{\textsf{\affiliation}}}\end{minipage}
\end{figure}

%%%%%%%%%%%%%%%%%%%%%%%%%%%%%%%%%%%%%%%%%%%%%%%%%%%%%%%%%%
%%%%%%%%%%%%%%%%%%%%%%%%%%%%%%%%%%%%%%%%%%%%%%%%%%%%%%%%%%
%%%%%%%%%%%%%%%%%%%%%%%%%%%%%%%%%%%%%%%%%%%%%%%%%%%%%%%%%%

%%%%%%%		 Main Text			%%%%%%% 

%	For Communications for Angewandte Chemie, please remove headlines for Introduction, Results and Discussion and Conclusion

\section*{Introduction}
\label{introduction}

Buckminsterfullerene, $\mathrm{C}_{60}$, the football shaped all-carbon molecule first found in a laboratory setting by Kroto and coworkers in 1985\cite{Kroto:1985}, is by now well-established to be widely present in different astronomical environments.\cite{Cami:2010,Campbell:2015} Likewise, Polycyclic aromatic hydrocarbons (PAHs), a related family of carbon-based molecules, have since long been proposed as carriers of infrared (IR) emission bands observed in interstellar environments (see reviews by Tielens\cite{Tielens:2008aa,Tielens:2013aa} and references therein). Yet, the first few specific PAH species were only recently unambiguously identified in the dark molecular cloud TMC-1 through radio astronomy observations.\cite{Mcguire:2021,Cernicharo:2021aa} In a context closer to earth, PAHs have been found to play a significant role in the formation of macroscopic soot particles, as first suggested by Haynes and Wagner\cite{HAYNES:1981,WAGNER:1979} and their formation mechanisms in flames are well studied.\cite{Richter:2000aa,Abdel-Shafy:2016aa} Similarly, the growth of PAH and fullerene clusters has been connected to the formation of circumstellar and interstellar dust grains.\cite{Tielens:2013aa} However, it has not yet been conclusively shown if PAHs and fullerenes are formed in bottom-up processes, top-down dito, or in both, nor is it fully known what roles such dust grains play in facilitating such mechanisms in astronomical environments.\cite{Berne:2012aa,Omont:2021aa,Herrero:2022aa}

The structures of pure clusters of fullerenes and PAHs have been studied thoroughly theoretically \cite{Doye:1996aa,Doye:2001aa,Rapacioli:2005aa,Rapacioli:2009aa,Chen:2014aa,Dontot:2019aa,Hansen:2022aa} as well in the laboratory.\cite{Martin:1993aa,Martin:1994aa,Hansen:1997aa,Manil:2003aa,Goulart:2017aa,Schouder:2019aa,Hansen:2022aa} C$_{60}$ has been shown to form decahedral and close-packed structures for clusters containing more than 13 molecules and icosahedral structures for smaller clusters using classical pairwise potentials and Monte Carlo methods.\cite{Doye:1996aa,Doye:2001aa} Rapacioli \emph{et al.}\ searched for the global energy minima for pure clusters of PAHs using similar classical methods, with species ranging in size from pyrene ($\mathrm{C}_{16}\mathrm{H}_{10}$) to circumcoronene ($\mathrm{C}_{54}\mathrm{H}_{18}$).\cite{Rapacioli:2005aa} They concluded that for clusters containing eight PAH molecules and smaller, the optimal formations are one-dimensional stacks. For larger clusters, the molecules start to favour so-called ``handshake'' structures containing two interacting bent stacks before approaching structures that mimic the bulk packing of benzene.\cite{Rapacioli:2005aa}

Compared to the many studies of pure $\mathrm{C}_{60}$ clusters and pure PAH clusters respectively, mixed clusters have not yet been studied as thoroughly. There are, however, a few exceptions that act as important precursors to this study: for instance, experiments have been performed investigating the behaviour of fullerene-PAH clusters when bombarded with energetic ions.\cite{PhysRevA.90.022713,Domaracka:2018} The aim of such studies has been to advance the understanding of ion-induced molecular growth mechanisms in cluster environments, and to investigate how such mechanisms and products differ depending on the cluster composition. It was shown that the mixing factor between the two molecular species indeed plays a significant role in the observed reactions and subsquent mass spectra.\cite{Domaracka:2018} The ion kinetic energy and mass of the projectile used in the collisions also had a great impact on the types of reaction products that were formed.\cite{PhysRevA.90.022713,Domaracka:2018} %Another example is (CIT.

\begin{figure}
\begin{center}
\includegraphics[width=\columnwidth]{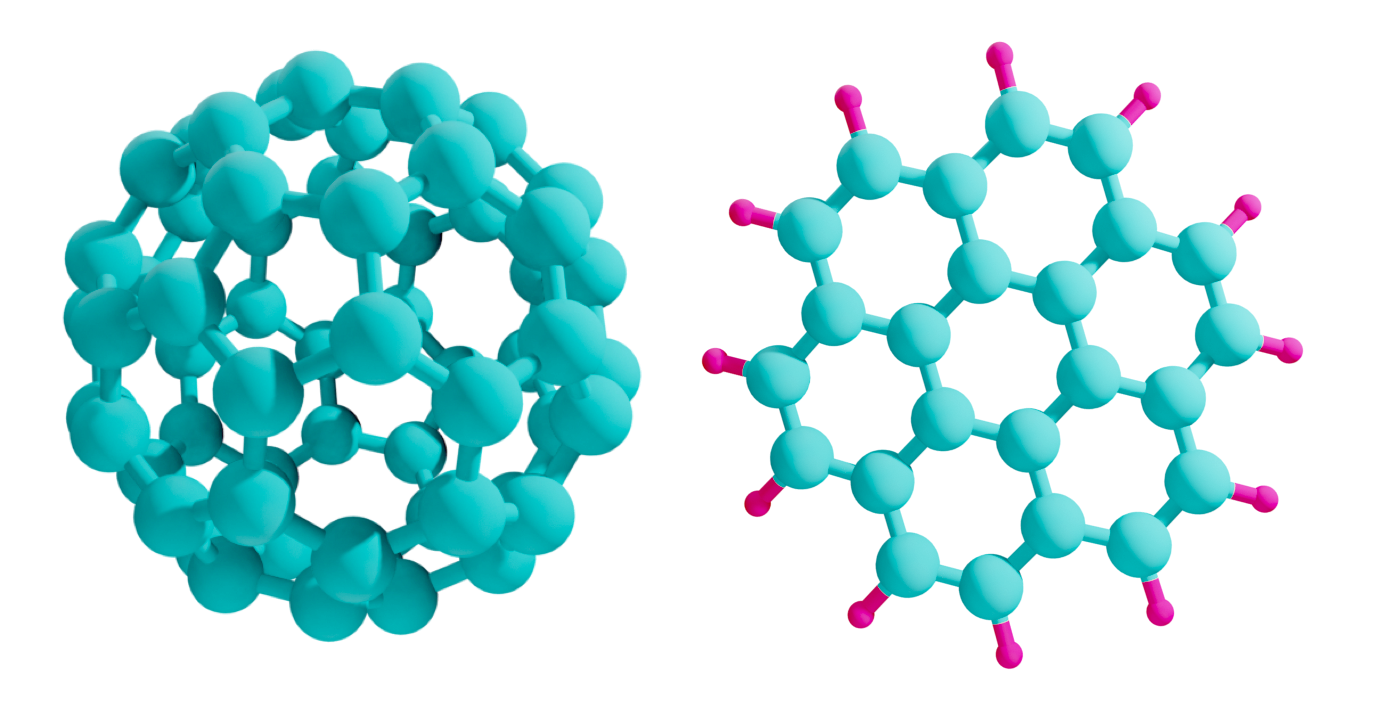}
\caption{Fullerene (C$_{60}$, left) and coronene (C$_{24}$H$_{12}$, right) molecules.}
\label{figcor}
\end{center}
\end{figure} 

Such experiments do not, however, provide detailed information on the structures of the clusters, be it for the precursors or products. In the experiments performed there,\cite{Domaracka:2018} it is neither possible to study single individual cluster sizes and their structures, nor to characterise the reaction products of the ion-cluster collisions beyond their mass-per-charge ratios.  Thus, much remains to be understood when it comes to the formation and optimal structures of mixed clusters of fullerenes and PAHs, and the impact that the structures have on the products formed by energetic processing. In this light, theoretical modelling offers a great potential of providing insight into the properties of these systems.

In this article, we present our findings from classical molecular dynamics studies of mixed clusters of $\mathrm{C}_{60}$ and $\mathrm{C}_{24}\mathrm{H}_{12}$ (coronene) molecules (see Figure \ref{figcor}).  Using randomised inputs, we probe clusters structures containing 5 and 13 mol\-e\-cules, respectively, of different mixtures. We observe and report on their structures and discuss trends in their relative stabilities.

\section*{Methods}

Given the large sizes of fullerene and coronene molecules, we use force field methods to calculate the total potential energies of clusters of such species, rather than \emph{ab initio} methods. The molecular dynamics (MD) software LAMMPS (Large-scale Atomic/Molecular Massively Parallel Simulator)\cite{Plimpton:1995aa,lammps_web} was used for all of the cluster simulations.

As initial centre-of-mass positions for the molecules in the cluster, we chose the optimised positions for pure C$_{60}$ clusters \cite{Doye:1996aa,Cluster_database}, and these were used for all runs. Which molecule was placed at what position was, however, randomised for each run, and as was the rotation of each molecule in three dimensions around its centre-of-mass. The distances between the molecules were slightly extended compared to the literature values for pure C$_{60}$ clusters to compensate for the fact that coronene has a larger diameter than C$_{60}$, thus ensuring that molecular overlap would not occur. 

The molecules were confined to a non-periodic, size-fixed square box with a side of 200 \AA. The force acting between the atoms, creating the H--H, C--C, and C--H bonds, was governed by the manybody AIREBO (Adaptive Intermolecular Reactive Empirical Bond Order) potential.\cite{Stuart:2000aa} This potential includes both intramolecular bonding interactions as well as intermolecular dispersion forces to accurately describe the structures of the molecular clusters.

Each configuration of molecules $(\mathrm{C}_{24}\mathrm{H}_{12})_{n}(\mathrm{C}_{60})_{N-n}$ for a fixed cluster size $N$ and mixture $n$ was run 10,000 times. In each run, the initial cluster geometry was optimised to obtain a preliminary minima structure. Following this, an MD simulation was carried out by heating the system to 500\,K and running for 1\,ps. This was done to introduce an additional degree of randomisation to the positions of the particles. Following this, the geometry was once again optimised and recorded to our ensemble of structures. With this method we aim to identify both putative lowest energy conformations as well as to gain insight into the breadth of different structures that may be formed experimentally in a gas aggregation cluster source.

We chose to run the simulation for sizes $N=5$ and $N=13$, with $n$ ranging from $n=0$ to $n=N$, thus also including non-mixed cluster variants. We chose the numbers 5 and 13 partly to match the ratios in previous experimental cluster work,\cite{Domaracka:2018} and partly because in the case of pure $\mathrm{C}_{60}$ clusters, $N=13$ is a magic number corresponding to the closure of the icosahedral shell.\cite{Doye:1996aa} Smaller clusters are computationally favoured over larger ones as well, both in terms of CPU time per simulation and the number of possible geometries available, therefore we limited ourselves to these manageable sizes. 

For the pure coronene clusters, we found stacks of PAHs for the smaller ($N=5$) clusters and the ``handshake structures'' for the slightly larger ($N=13$) clusters, as predicted by Rapacioli \emph{et al.}\cite{Rapacioli:2005aa} While they study equivalent systems of PAHs, their approach for minimisation differs significantly from ours. The fact that our model nevertheless was able to find these known cluster structures strengthens the credibility of the generalisability of our method. Likewise, the ground state fullerene structures we identify agree with previous results.\cite{Doye:1996aa}

\begin{figure}
\begin{center}
\includegraphics[width=\columnwidth]{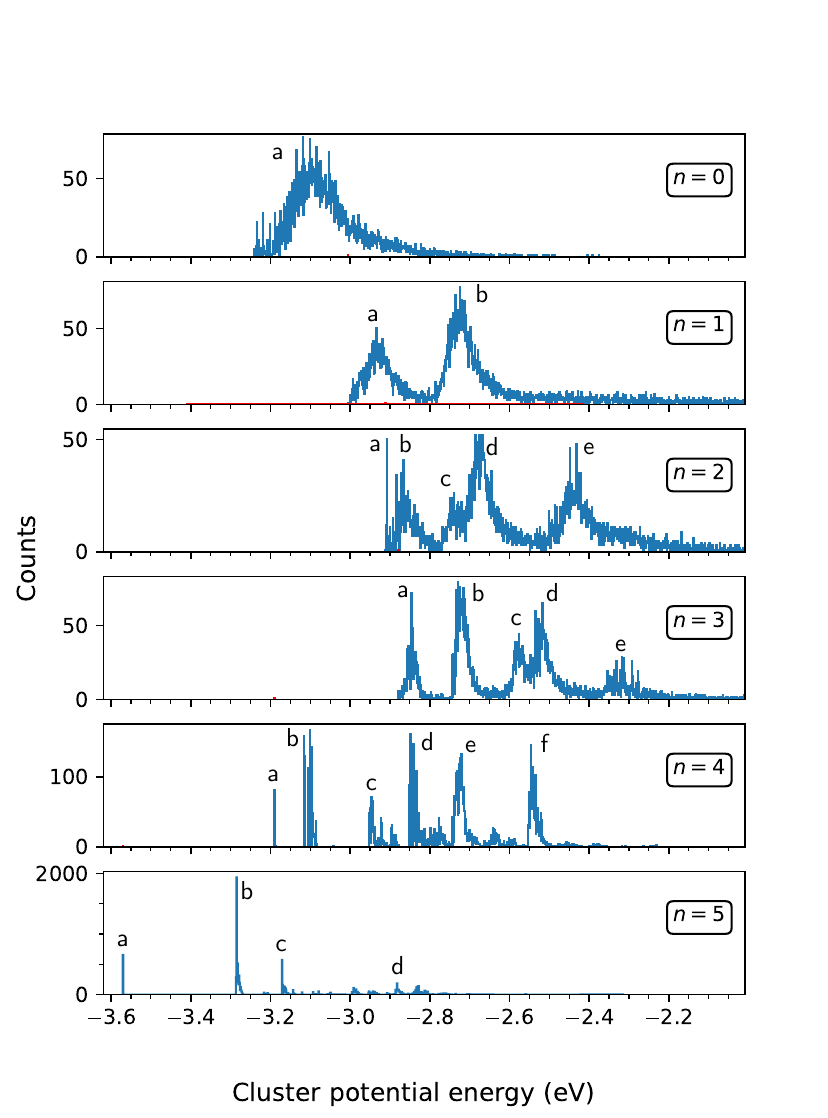}
\caption{Distribution of potential energies, $E_p$, for pure and mixed coronene and fullerene clusters containing five molecules, $($C$_{24}$H$_{12})_n($C$_{60})_{5-n}$ $(n=0-5)$. The greater number of coronene molecules present in the cluster, the more diverse the energy distribution. Images of typical cluster structures with potential energies corresponding to the markings in each plot (a, b, c ...) can be found in the supplemental material. Note the differences in vertical scale between the panels.}

\label{fig1}
\end{center}
\end{figure} 

\section*{Results and Discussion}
\label{results_discussion}

The potential energy, $E_p$, for a $(\mathrm{C}_{24}\mathrm{H}_{12})_n(\mathrm{C}_{60})_{N-n}$ cluster of size $N$ with $n$ coronene molecules and $N-n$ fullerenes was calculated as 

\begin{equation}
\label{eq1}
E_p= E_p^{tot}-nE_c-(N-n)E_f
\end{equation}
where $E_p^{tot}$ is the potential energy of the whole system, $E_c$ the potential energy of an isolated coronene molecule and $E_f$ the potential energy of an isolated fullerene molecule.

\begin{figure}
\begin{center}
\includegraphics[width=\columnwidth]{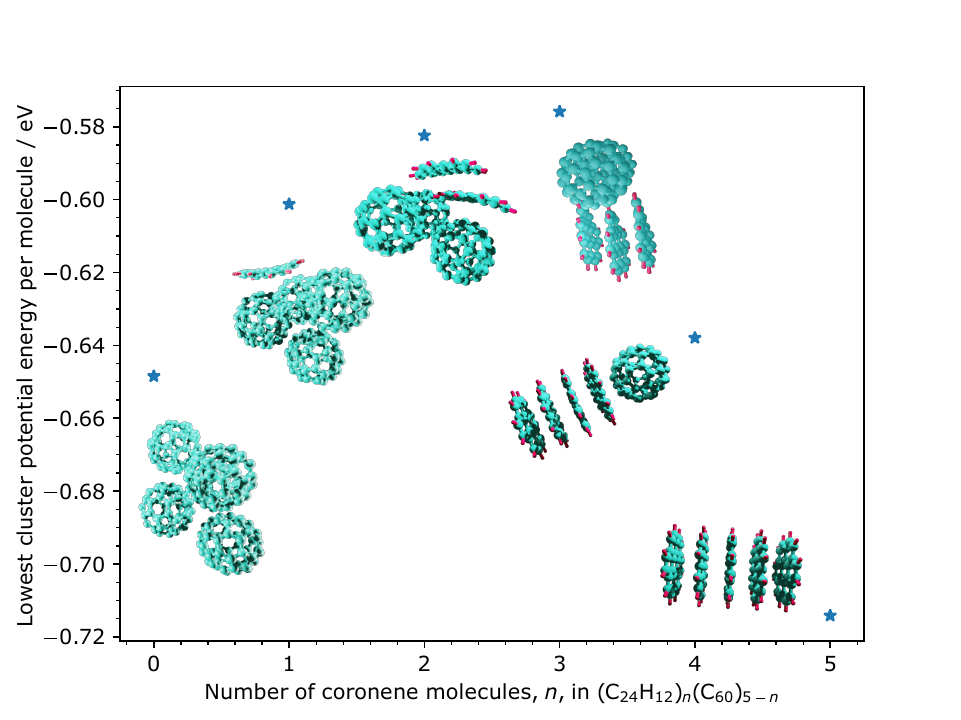}
\caption{The potential energies per molecule ($E_p/N$) and visualisations of the lowest-energy $N=5$ clusters, $($C$_{24}$H$_{12})_n($C$_{60})_{5-n}$ $(n=0-5)$.}
\label{fig2}
\end{center}
\end{figure}

\subsection*{$N=5$ clusters}

Figure \ref{fig1} shows the distributions of potential energies ($E_p$) for $(\mathrm{C}_{24}\mathrm{H}_{12})_n(\mathrm{C}_{60})_{5-n}$ clusters $(n=0,1,2,3,4,5)$, as calculated using eq.\ (\ref{eq1}). The peaks in each panel are labelled (a, b, c, etc.), where each label represents a family of structures with similar energies. Representative structures from these distinct structure families can be found in the supporting information.

In the case when $n=1$, the clusters tend to form two distinct structure families (labelled a and b, respectively), one with lower energy and one with slightly higher, as seen in the top panel of figure \ref{fig1}. Upon further investigation, we find that the left peak (a) corresponds to the $\mathrm{C}_{60}$ molecules forming a tetrahedral structure, whereas the right peak (b) represents a planar $\mathrm{C}_{60}$ distribution. This is consistent with the stability order for pure $(\mathrm{C}_{60})_{4}$ clusters.\cite{Doye:1996aa} The width of each distribution then comes from the lone coronene molecule occupying different positions on the surface of the fullerene cluster. In a similar fashion, the most stable $n=2$ clusters consist of the ground state triangular $(\mathrm{C}_{60})_{3}$ cluster structure interacting with a stack of two coronene molecules, rather than a more even mixing of the two species. Moving to $n= 3$ and $4$, the lowest energy structures (a) correspond to clusters with a single PAH stack, while clusters with two or more stacks (or no stacks) are higher in energy. Effectively, this means that the coronene and $\mathrm{C}_{60}$ molecules tend to separate within the cluster, forming two sub-clusters of coronene and fullerene molecules that each are similar to their isolated versions in structure. We also see that the lower energy clusters in general are more compactly bound than the higher energy ones.

The increasing number of peaks for increasing $n$ can be explained by the fact that fullerenes are approximately spherical and there are few distinct ways in which fullerene mol\-e\-cules can sort themselves into a cluster of a given size. For the planar coronene molecules on the other hand, the number of possible arrangements increases significantly as more coronene molecules are introduced into the clusters. For example, individual coronene molecules may be offset parallel to their planes and even interact edge-on as apposed to the more typically more energetically favourable face-on configuration. As we will see though, the edge-on interactions becomes increasingly important as the cluster sizes increase and stacks of coronene interact with their surroundings.

The cluster structures for each $n$ with the lowest energy are shown in figure \ref{fig2}, along with their corresponding potential energies per molecule. Here, it is clear that for clusters of size $N=5$, the lowest energy occurs for $n=5$, that is, for a pure coronene cluster where all molecules are all aligned in a single linear stack. This reflects the significantly larger pair-wise interaction energy between neighbouring coronene molecules ($-0.86$\,eV with the AIREBO potential) compared to that of two interacting fullerenes ($-0.38$\,eV). However, the potential energy per molecule is similar for pure coronene ($-0.71$\,eV) and C$_{60}$ ($-0.63$\,eV) clusters as the monomers in the latter case interact strongly with more than one nearest neighbour. The clusters with the largest degree of mixing ($n=2$ and 3) are the least stable ones.

\begin{figure}
\begin{center}
\includegraphics[width=\columnwidth]{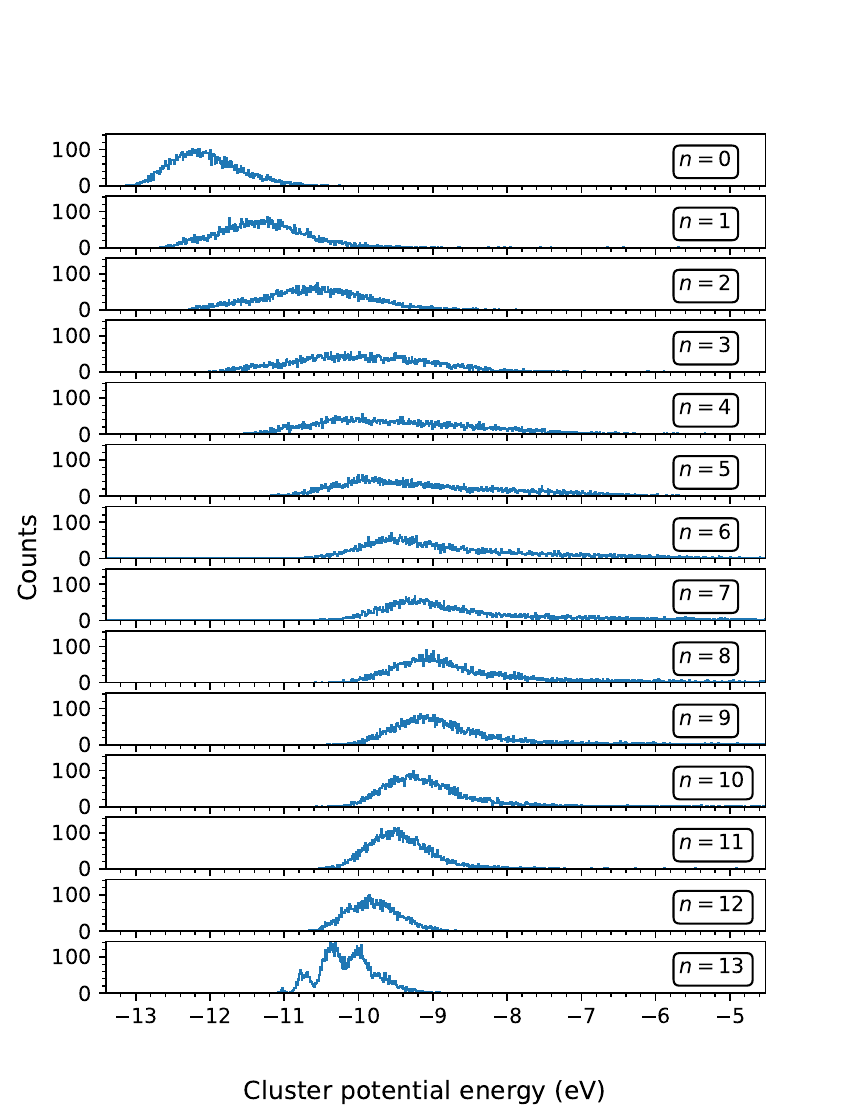}
\caption{Distribution of potential energies ($E_p$) for pure and mixed coronene and fullerene clusters containing thirteen molecules, $($C$_{24}$H$_{12})_n($C$_{60})_{13-n}$ $(n=0-13)$.}
\label{fig3}
\end{center}
\end{figure}

\subsection*{$N=13$ clusters}

The potential energy distributions for larger clusters of size $N=13$ are shown in Figure \ref{fig3}. In contrast to the distributions for smaller clusters (Fig.\ 2), the energy distributions for each $n$ are centred around a single continuous hump. Unlike for $N=5$, we therefore cannot simply identify distinct structure families for each $n$ based on their energies alone. We attribute this to the greater complexity of the potential energy surface of the larger clusters. However, that the lower energy clusters are more compactly bound than their high energy equivalents seems to also be true in this case of $N=13$.

\begin{figure}
\begin{center}
\includegraphics[width=\columnwidth]{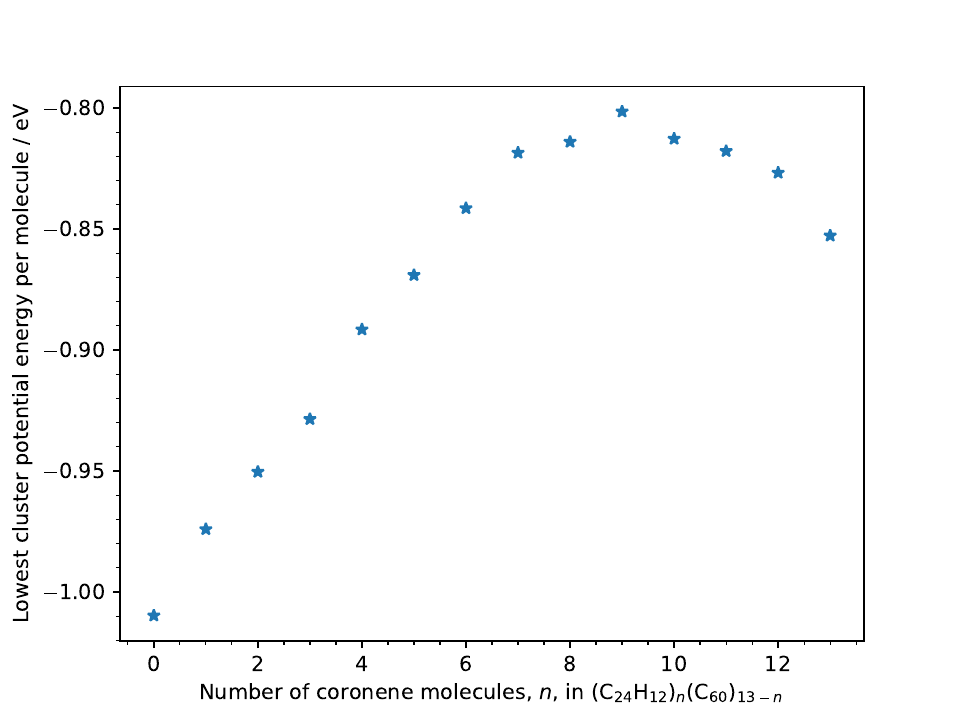}
\caption{The potential energies per molecule ($E_p/N$) for the lowest-energy $N=13$ clusters, $($C$_{24}$H$_{12})_n($C$_{60})_{13-n}$ $(n=0-13)$.}
\label{fig4}
\end{center}
\end{figure} 

\begin{figure}
\begin{center}
\includegraphics[width=\columnwidth]{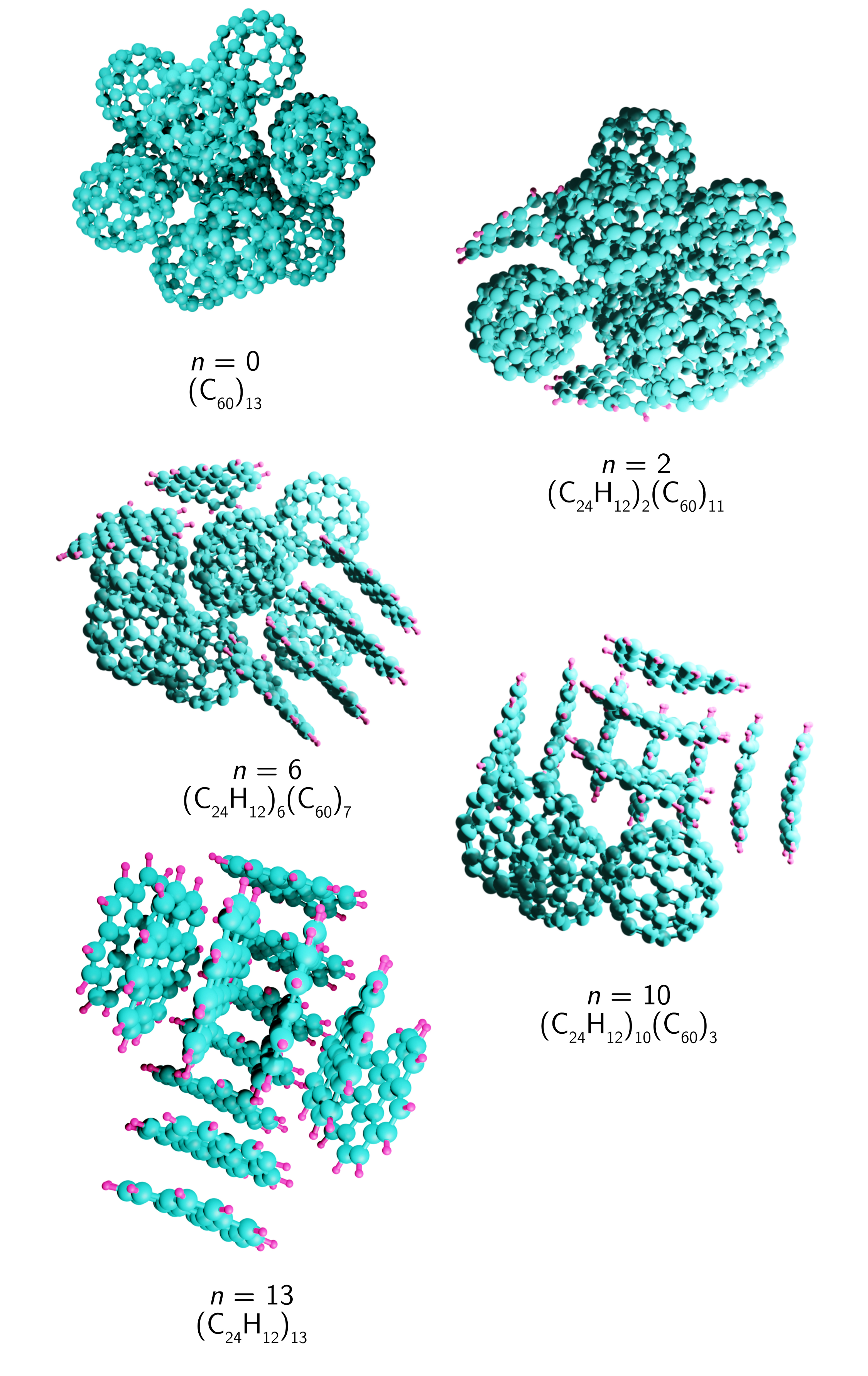}
\caption{Selection of lowest energy structures for clusters of size $N=13$ ($n=0, 2, 6, 10, 13$). Additional structures can be found in the supporting information.}
\label{fig5}
\end{center}
\end{figure}

The proposed global energy minima of each mixture $n$ for cluster size $N=13$ are shown in Figure \ref{fig4} and a selection ($n=0,6,10,13$) of the structures  in Figure \ref{fig5}. In contrast to the case of $N=5$, the $n=0$ cluster (pure fullerene cluster) has a lower potential energy per molecule than that of $n=13$ (pure coronene cluster). This can be explained intuitively by the fact that it is easier to pack spheres than planar shaped molecules compactly together in three dimensions. The potential energy of about $-1$\,eV per molecule is nearly three times higher than the pairwise interaction energy, showing that the fullerenes indeed interact with multiple neighbours in this icosahedral cluster geometry. At the other extreme, for the pure coronene clusters ($n=13$) the well-known handshake structure is preferred over the single stack. At this size this structure consists of two evenly sized  stacks (6 and 7 molecules, respectively) that cusp each other along the edges of the molecules (and stacks).

For intermediate mixtures we observe a general trend where the fullerene and PAH clusters form separate units. For $n\lesssim 6$, the C$_{60}$ molecules form a cluster very similar to pure clusters of the same size. This core is then surrounded by individual coronene molecules that pack tightly along its surface. An example of this can be seen in the $(\mathrm{C}_{24}\mathrm{H}_{12})_{2}(\mathrm{C}_{60})_{11}$ structure ($n=2$) in Figure \ref{fig5}. Around $n=6$, structures where several of the PAH molecules form a single stack begin to dominate the lowest energy geometries. The $(\mathrm{C}_{24}\mathrm{H}_{12})_{6}(\mathrm{C}_{60})_{7}$ structure in Figure \ref{fig5} has a single stack containing four coronene molecules that binds to the fullerene cluster. The remaining two coronene molecules bind directly to the same. Finally, around $n=10$ we start observing mixed clusters with twin coronene stacks as the most tightly bound systems. The $(\mathrm{C}_{24}\mathrm{H}_{12})_{10}(\mathrm{C}_{60})_{3}$ cluster with the lowest potential energy found in our calculations consists of two stacks with seven and three molecules, respectively, with the remaining three fullerenes forming a triangular substructure (Figure \ref{fig5}). This trend then continues to the pure coronene clusters.

As is the case for the smaller clusters, the least stable mixed systems are found when half to two-thirds of the cluster constituents are coronene molecules. Again, this points towards an inefficient mixing of the two molecular types which is likely caused by their different geometries.

\section*{Summary and Outlook}
\label{conclusion}

We have studied the cluster formation of two sizes of mixed clusters of coronene and $\mathrm{C}_{60}$ molecules, 5 and 13 molecules in total, respectively, using the classical manybody AIREBO force field. Our method gave the expected cluster structures for pure clusters based on previous studies,\cite{Doye:1996aa,Rapacioli:2005aa,Rapacioli:2009aa} and new results for the mixed systems. For the smaller size, the pure coronene cluster display the highest stability, while for the larger cluster size the pure fullerene cluster is the most stable. We attribute this to the geometries of the molecules as interacting spheres that can be packed into three-dimensional structures more efficiently than discs. As for the mixed clusters, a general trend is that the two molecule types tend to form separate sub-units and display a rather poor mixing. This behaviour is somewhat less apparent though in the larger clusters in cases when the fullerenes outnumber the coronene molecules, in which case the latter tend to adhere as individual flakes to the surface of the C$_{60}$ clusters. Another trend that is clearly seen for all cluster combinations is that they tend to prefer more compactly bound structures over more spatially spread-out ones.

Here, we have focused on the structure and stabilities of mixed PAH-fullerene clusters. Our interest in these systems is based, in part, on their abundance in astronomical environments and the potential roles they may play in interstellar chemistry. Experimental studies where keV ions collide with mixed clusters of these types have shown that a wealth of new molecular species can be produced, and that the specific products formed can vary significantly depending on ionic species and energy of the projectile, as well as the composition of the mixed cluster precursors.\cite{Domaracka:2018,PhysRevA.90.022713} Our simulations show that the coronene and C$_{60}$ molecules mix poorly, which will most certainly influence the types of reactions that can occur in these types of collisions. This could suggest that reactions between fragments from the same precursor species would be favoured over mixed reaction products, but it remains speculative at this point and additional modelling will be necessary to better understand these reactions.

\section*{Acknowledgements}
HZ and MG acknowledge financial support from the Swedish Research Council (contracts 2020-03437 and 2020-03104) and from the project grant “Probing charge- and mass-transfer reactions on the atomic level” (2018.0028) from the Knut and Alice Wallenberg Foundation. This publication is based upon work from COST Actions CA18212 -- Molecular Dynamics in the GAS phase (MD-GAS), and CA21101 -- Confined Molecular Systems: from the new generation of materials to the stars (COSY), supported by COST (European Cooperation in Science and Technology).
%Acknowledgements text.

\section*{Conflict of Interest}

There are no conflicts to declare.

%%%%%%%%%%%%%%%%%%%%%%%%%%%%%%%%%%%%%%%%%%%%%%%%%%%%%%%%%%
%%%%%%%%%%%%%%%%%%%%%%%%%%%%%%%%%%%%%%%%%%%%%%%%%%%%%%%%%%
%%%%%%%%%%%%%%%%%%%%%%%%%%%%%%%%%%%%%%%%%%%%%%%%%%%%%%%%%%
\begin{shaded}
\noindent\textsf{\textbf{Keywords:} \keywords} 
\end{shaded}
%%%%%%%%%%%%%%%%%%%%%%%%%%%%%%%%%%%%%%%%%%%%%%%%%%%%%%%%%%
%%%%%%%%%%%%%%%%%%%%%%%%%%%%%%%%%%%%%%%%%%%%%%%%%%%%%%%%%%
%%%%%%%%%%%%%%%%%%%%%%%%%%%%%%%%%%%%%%%%%%%%%%%%%%%%%%%%%%

%%%%%%%		References			%%%%%%% 

\setlength{\bibsep}{0.0cm}
\bibliographystyle{Wiley-chemistry}
\bibliography{references}

%%%%%%%%%%%%%%%%%%%%%%%%%%%%%%%%%%%%%%%%%%%%%%%%%%%%%%%%%%
%%%%%%%%%%%%%%%%%%%%%%%%%%%%%%%%%%%%%%%%%%%%%%%%%%%%%%%%%%
%%%%%%%%%%%%%%%%%%%%%%%%%%%%%%%%%%%%%%%%%%%%%%%%%%%%%%%%%%

\end{document}